    \documentstyle[epsf]{article} \catcode`\@=11
    \def\section{\@startsection{section}{1}{\z@}%
    {-3.5ex plus -1ex minus -.5ex}{1.5ex plus.3ex}{\bf }}
    \def\subsection{\@startsection{subsection}{1}{\z@}%
    {-3.5ex plus-1ex minus-.5ex}{1.5ex plus.3ex}{\bf }} 
    
\begin{document}
{\Large\bf
A numerical study \\ of wave-function and matrix-element 
statistics \\ in the Anderson model of localization
    }\vspace{.4cm}\newline{\bf   
V. Uski$^1$, B. Mehlig$^2$, and R.A. R\"omer$^1$
    }\vspace{.4cm}\newline\small
$^1$Institut f\"ur Physik, Technische Universit\"at, D-09107 Chemnitz, 
Germany\\ 
$^2$Theoretical Physics, University of Oxford, 1 Keble Road, OX1 3NP, UK
    \vspace{.2cm}\newline 
Received 6 October 1998, revised 16 October 1998, accepted in final form
23 October 1998 by M. Schreiber
    \vspace{.4cm}\newline\begin{minipage}[h]{\textwidth}\baselineskip=10pt
    {\bf  Abstract.}
    We have calculated wave functions and matrix elements
    of the dipole operator in the two- and three-dimensional Anderson 
    model of localization and have studied their statistical 
    properties in the limit of weak disorder. 
    In particular, we have considered two cases. 
    First, we have studied the fluctuations
    as an external Aharonov-Bohm flux is varied.
    Second, we have considered the influence
    of incipient localization. In both cases,
    the statistical properties of the eigenfunctions
    are non-trivial, in that the
    joint probability distribution function
    of eigenvalues and eigenvectors does no longer
    factorize. We report on detailed comparisons
    with analytical results, obtained within
    the non-linear sigma model and/or
    the semiclassical approach.
    \end{minipage}\vspace{.4cm} \newline {\bf  Keywords:}
Disorder, semiclassical approximation, wave function statistics
    \newline\vspace{.2cm} \normalsize

\section{Introduction}
Disordered quantum systems  exhibit irregular fluctuations of eigenvalues, 
wave functions and matrix elements. 
The statistical properties of wave functions  and matrix elements
are of particular interest. 
Both are of direct
experimental relevance. Fluctuations
of wave-function amplitudes determine 
the fluctuations of the conductance through
quantum dots. The effect
of an external perturbation is described
by matrix elements of the perturbing operator
in the eigenstates of the system.

In the metallic regime (which is characterized by a large conductance
$g \gg 1$), wave-function and matrix-element fluctuations are described
by random matrix theory (RMT) \cite{meh67,efe97}. In Dyson's ensembles
\cite{meh67} such fluctuations are particularly simple since
the joint probability distribution function
of eigenvector components and eigenvalues factorizes \cite{por65}
and the statistical properties of the eigenvectors
are determined by the invariance properties of
the random matrix ensembles. One finds that
the wave-function amplitudes are distributed according to
the Porter--Thomas distribution \cite{por65}.
Non-diagonal matrix elements of an observable
$\widehat{A}$ are
Gaussian distributed \cite{por65} around
zero with variance $\sigma_{\rm off}^2$. 
Diagonal matrix elements are also Gaussian distributed,
with variance
$\sigma_{\rm diag}^2 = (\beta/2) \sigma_{\rm off}^2$
where $\beta = 1$ in the Gaussian orthogonal ensemble (GOE)
of random matrices and $\beta = 2,4$ in the
Gaussian unitary (GUE) and symplectic ensembles.
The variance of non-diagonal matrix elements
is essentially given by a time integral
of a classical autocorrelation function
\cite{wilk87}
and does not depend on the symmetry properties.

Of particular interest are those cases where
the fluctuations of eigenvalues and eigenvectors are no longer independent.
In the following we analyse two such situations. First,
we consider the effect of an Aharonov-Bohm flux which
breaks time-reversal invariance and
drives a transition from the GOE to the GUE.
The statistical properties of diagonal matrix
elements in the transition regime
between GOE and GUE were calculated
in \cite{meh98,tan94,kea98},
those of level velocities in \cite{tan94,bra94,usk98}.
The statistical
properties of wave functions
in the transition regime
were derived in \cite{fal94}.
Here we compare these predictions
with numerical results obtained
for the Anderson model.
Second, we study the influence of increased disorder
on the statistical properties of wave functions.
The question how the distribution
of wave-function amplitudes deviates
from the RMT predictions at
smaller conductances $g$ has
recently been discussed in \cite{mir93}, within the 
framework of the non-linear sigma model
\cite{fal95a,fyo95,fal95b}, and 
using a direct optimal fluctuation
method \cite{smol97}.
Here we compare  our numerical
results for distributions
of wave-function amplitudes
in the $d=3$ dimensional Anderson model
with the perturbative results
of \cite{fyo95}.

\section{The Anderson model of localization}
We consider the Anderson model of localization  \cite{and58}
in $d=2$ and $d=3$ dimensions which is defined
by the tight-binding Hamiltonian on a
square or cubic lattice with $N$ sites
and unit lattice spacing
\begin{equation}\label{eq:H}
  \widehat H= \sum_n|n\rangle\epsilon_n\langle n| + \sum_{n\neq
    m}|n\rangle t_{nm} \langle m|\,,
\end{equation}
and $|n\rangle$ represent the Wannier state at site $n$.
The on-site potential $\epsilon_n$ is
assumed to be random and
taken to be uniformly distributed between $-W/2$ and $W/2$. 
The hopping parameters $t_{nm}$ connect only nearest-neighbour
sites (and $t=1$).
In the presence of an Aharonov Bohm-flux $\varphi$, 
the hopping parameters acquire an
additional phase. The flux $\varphi$ is measured in units of the flux quantum
$\varphi_0=hc/e$ and we define
$\phi = \varphi/\varphi_0$. The presence of
an Aharonov-Bohm flux $\phi$ breaks time-reversal invariance.

We have determined the eigenvalues $E_\alpha$
and the eigenfunctions $\psi_\alpha(\bf{r})$
using a modified Lanczos algorithm \cite{cul85}.
The statistical properties of
eigenvalues, eigenfunctions
and matrix elements \cite{note}
$A_{\alpha\beta} = \langle \psi_\alpha | \widehat A |\psi_\beta\rangle$
of an observable $\widehat A$ depend on the strength
of the disorder potential. In the metallic regime
(which is characterized by a large conductance $g$)
these fluctuations are described by RMT
on energy scales smaller
than the Thouless energy $E_{\rm D} = g \Delta$,
where $\Delta$ is the mean level spacing. This
is the regime we consider in Sec.\ 3, while
Sec.\ 4 deals with $g^{-1}$ corrections
to the distribution of wave-function
amplitudes.

\section{The GOE to GUE transition}
In this section, 
we discuss numerical results 
for the fluctuations of
matrix elements, level velocities and wave functions
in the transition region between GOE and GUE.
We have calculated the smoothed variances
\begin{equation}
C_{\mathrm v}(\epsilon,\phi) = 
\langle |\widetilde{d}_{\mathrm v}(E,\phi;\epsilon)|^2\rangle_E
\hspace*{1cm}
C_{\mathrm m}(\epsilon,\phi) = 
\langle |\widetilde{d}_{\mathrm m}(E,\phi;\epsilon)|^2\rangle_E
\end{equation}
where $\widetilde{d}_{\mathrm{v,m}}(E,\phi;\epsilon)$ are
the fluctuating parts of the densities
\begin{eqnarray}
d_{\mathrm v}(E,\phi;\epsilon)&=&\sum_\alpha 
\frac{\partial E_\alpha}{\partial\phi}\,
\delta_\epsilon[E-E_\alpha(\phi)]\\
d_{\mathrm m}(E,\phi;\epsilon)&=&\sum_\alpha A_{\alpha\alpha}\,\,
\delta_\epsilon[E-E_\alpha(\phi)]
\end{eqnarray}
with $\delta_\epsilon(E)=
(\sqrt{2\pi}\epsilon)^{-1}\exp(-E^2/2\epsilon^2)$. 
It is assumed that
$\langle A_{\alpha\alpha}\rangle  = 
\langle \partial E_\alpha/\partial\phi\rangle=0$.
$C_{\mathrm m}$ was calculated
exactly in \cite{meh98}. 
Here
we compare with semiclassical
expressions for $C_{\mathrm m}$ \cite{meh98,kea98}
and $C_{\mathrm v}$ \cite{usk98} 
obtained within the diagonal approximation
and valid for $ \Delta < \epsilon < g\Delta$ \cite{note2}
(see also \cite{bra94}).
In the case of $C_{\mathrm m}$ we considered the matrix elements of
the dipole operator $\widehat{A}=\hat{x}$.
\begin{figure}[t]
\centerline{\epsfysize=5.5cm\epsfbox{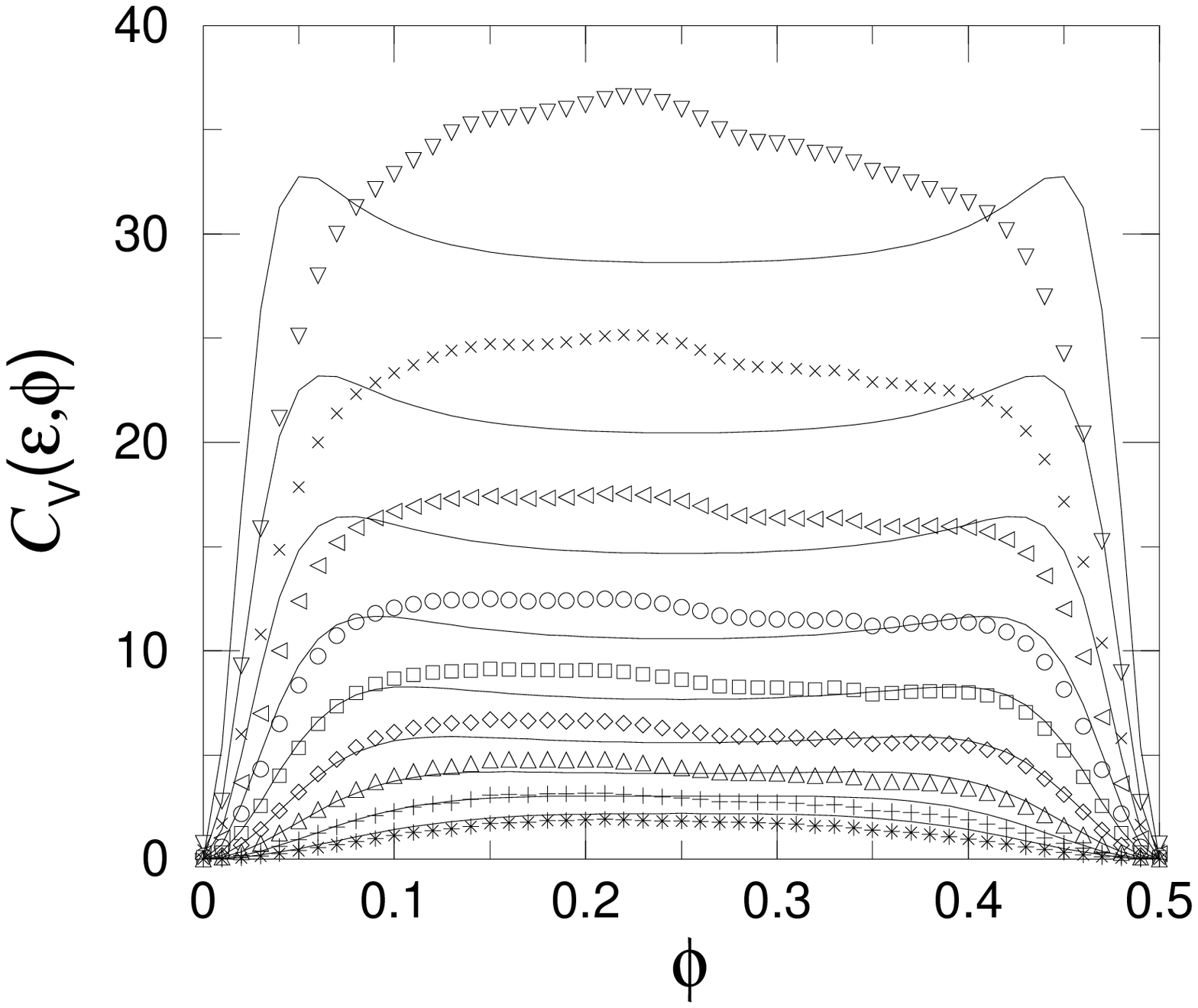}
\hfill
 \epsfysize=5.5cm\epsfbox{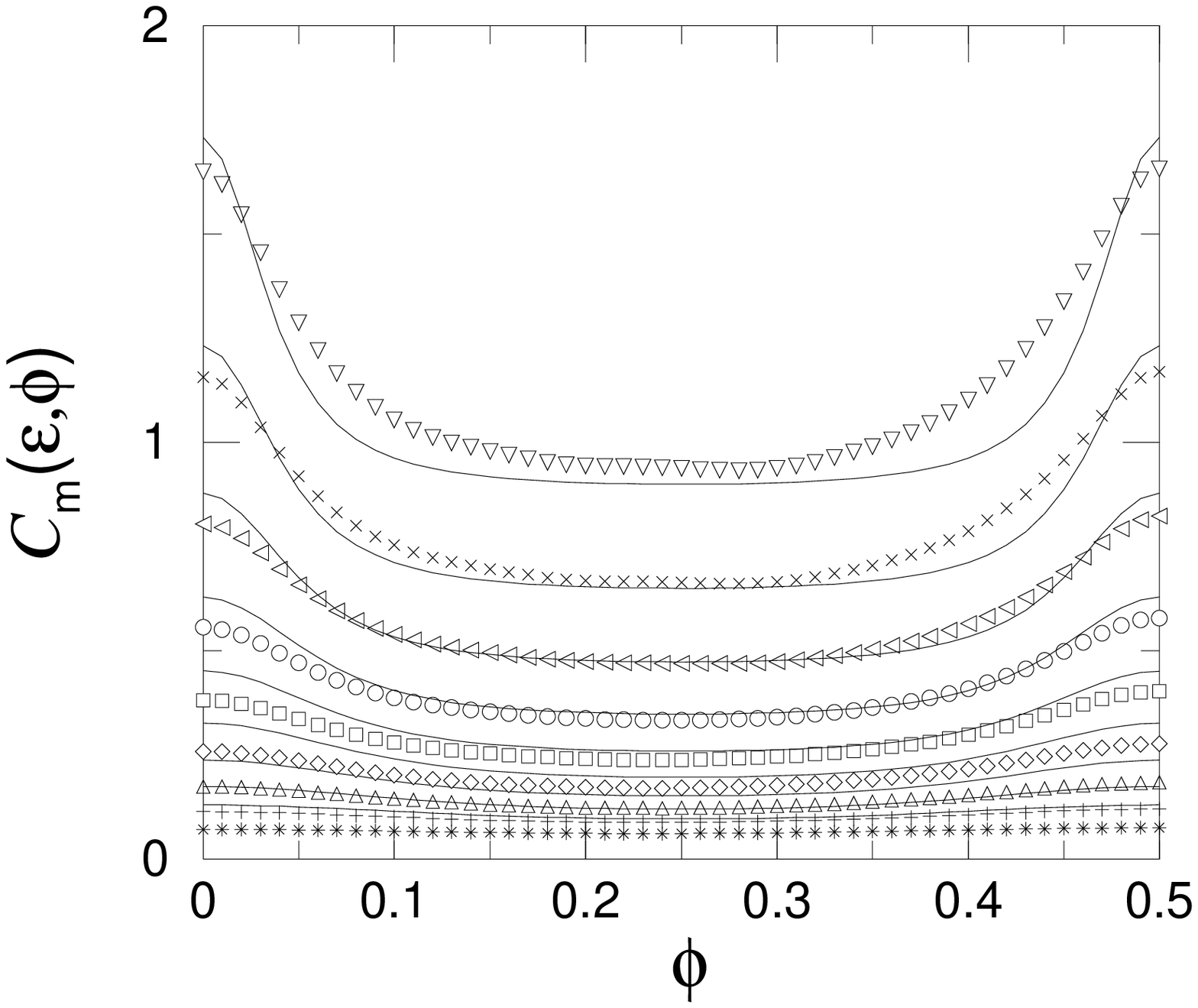}
}
\caption{\label{fig:cv:cm}\protect\small Velocities
and matrix elements in the transition
regime between GOE and GUE for the $27\times 27$
 Anderson model with $W=1.7$ (symbols), 
compared to the corresponding
semiclassical expressions (lines).
The numerical values of $\epsilon$ are $0.158$ 
($\bigtriangledown$), $0.224$ ($\times$), $0.316$ ($\lhd$), $0.447$ 
  ($\circ$), $0.631$ ($\Box$), $0.891$
  ($\Diamond$), $1.26$ ($\bigtriangleup$), $1.78$ ($+$), $2.51$
  ($\ast$).}
\end{figure}
The numerical data were obtained
by averaging over 69 realisations of disorder in
the $27\times 27$ Anderson model with $W=1.7$
and over all states in the energy interval
$[-3.4,-1.9]$. The two-dimensional case was considered in order to be
able to obtain a good numerical accuracy at each flux value in a 
tolerable computing time. In general we observe good agreement.
For a more detailed discussion of the results see \cite{usk98}.

In Fig.\ 2(a)  we show corresponding results
for the distribution function 
of wave-function amplitudes which is
defined as follows
\begin{figure}
\centerline{\epsfysize=5.5cm\epsfbox{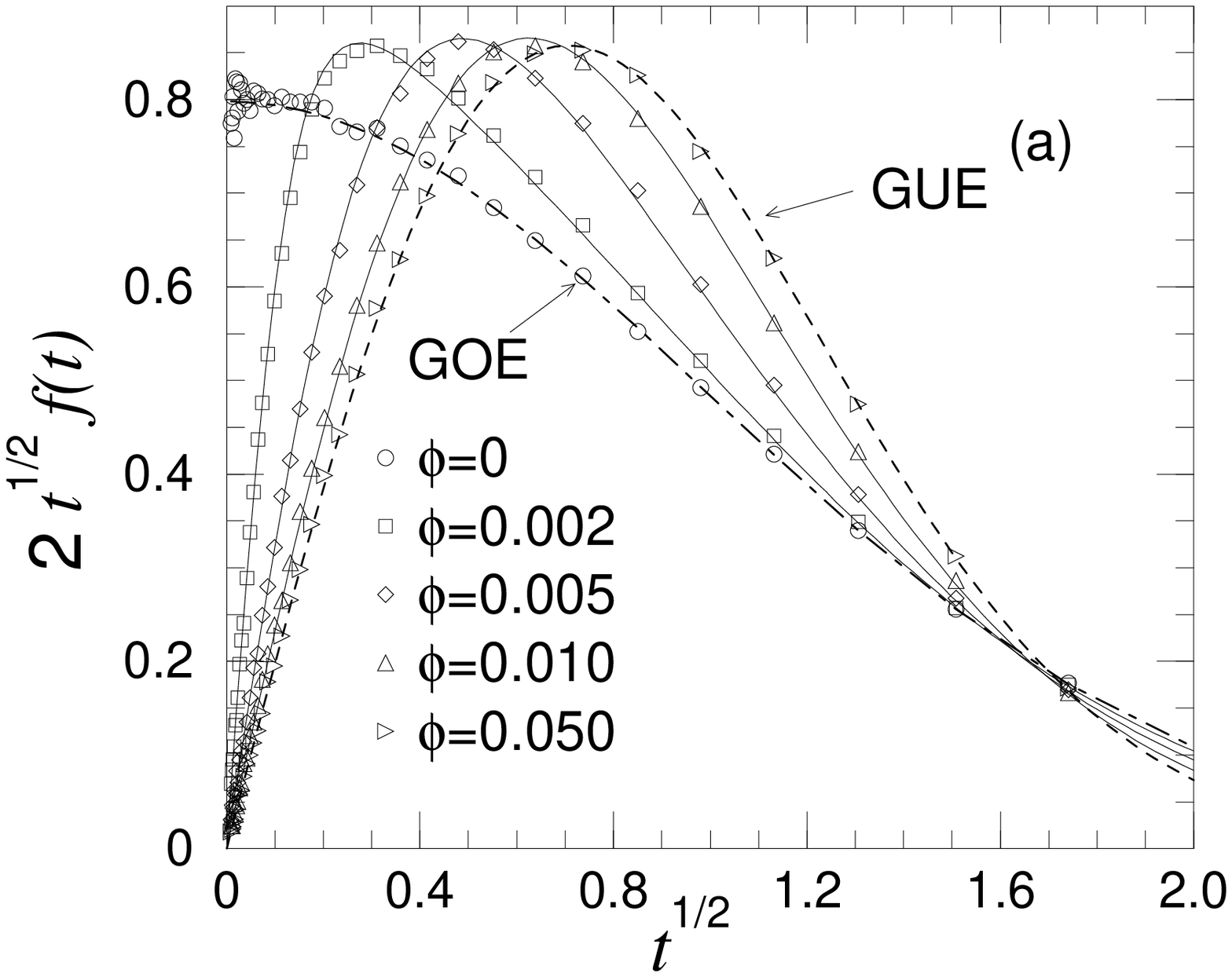}
\hfill\epsfysize=5.5cm\epsfbox{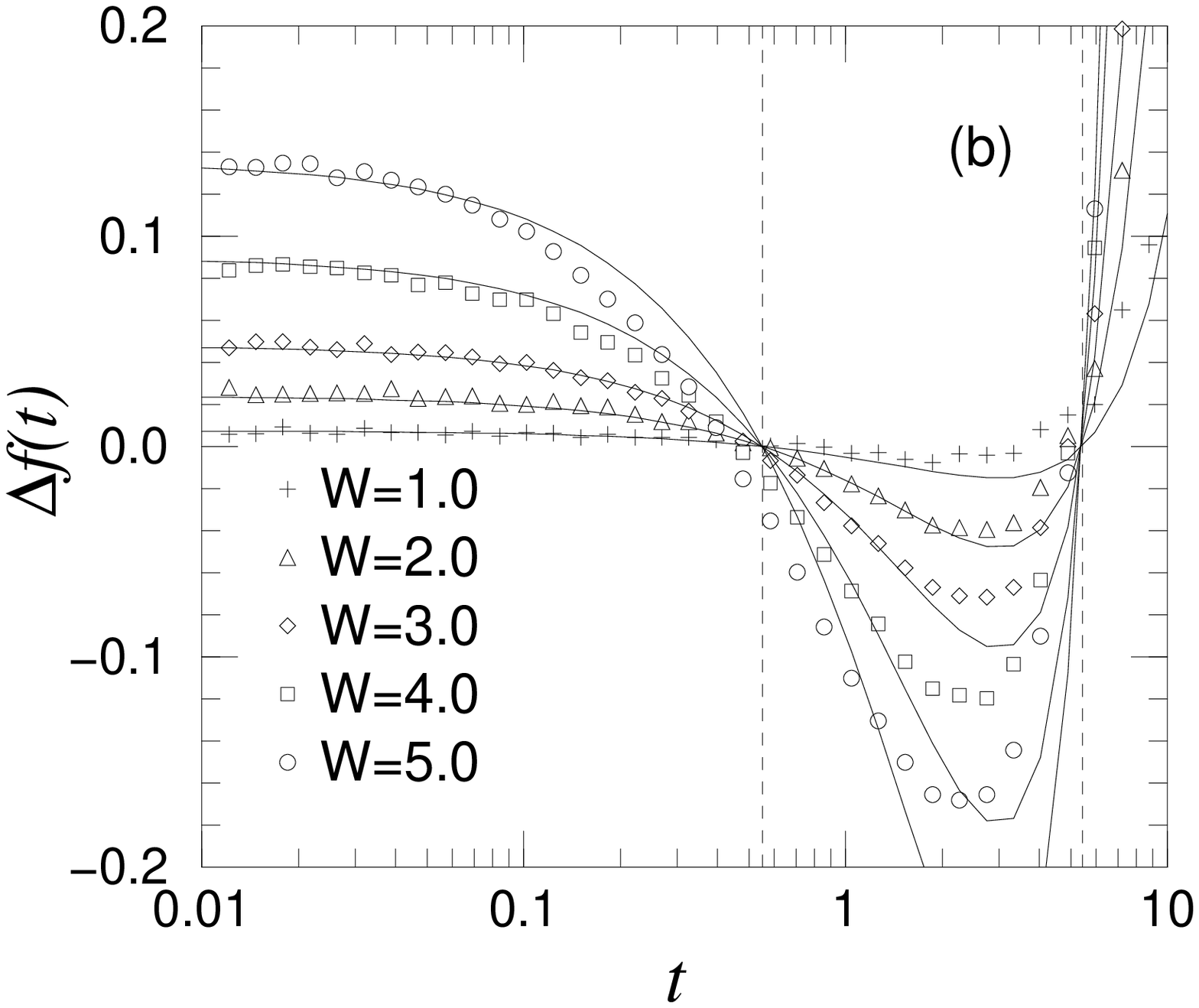}}
\caption{\label{fig:co}
\protect\small 
Wave-function statistics in the $13\times 13 \times 13$ 
Anderson model. (a) GOE to GUE transition
of the distribution function
in the metallic regime.
The Porter-Thomas distributions
are shown as dashed lines.
The predictions of \protect\cite{fal94}
are show as solid lines.
(b) Deviations $\Delta f(t)$
from the Porter--Thomas distribution
for the orthogonal case, and
several values of disorder
strength, $W=1,2,3,4$ and $5$. The results are
averaged over the energy interval $[-1.7,-1.4]$.
The dashed vertical lines denote the zeros of the first order correction term
in Eq.\ (\protect\ref{eq:co}). The solid
lines are fits according to Eq.\ (\protect\ref{eq:co}).
}
\end{figure}
\begin{equation}
f(t)=\Delta\left\langle\sum_\alpha\delta(t-|\psi_\alpha({\bf r})|^2N)\,
\delta(E-E_\alpha)\right\rangle\,.
\end{equation}
Within RMT one obtains for $f(t)$ in the limiting cases
of GOE and GUE
\begin{eqnarray}
f_{\mathrm PT}^{\rm GOE}(t)&=&{1\over \sqrt{2\pi t}}\exp(-t/2)\,,\\
f_{\mathrm PT}^{\rm GUE}(t)&=&\exp(-t)
\end{eqnarray}
(Porter--Thomas distributions \cite{por65}).
An expression in the transition
region between GOE and GUE was derived in \cite{fal94}.
We have computed $f(t)$ numerically for
several values of $\phi$ in the $13 \times 13 \times 13$
Anderson model at $W=0.5$ (in the metallic regime).  
Our results are shown in Fig.\ 2(a), where 
we have plotted $2\sqrt{t} f(t)$ versus $\sqrt{t}$.
The results are in good agreement with the analytical formulae
(6), (7) and those given in \cite{fal94}.

It was predicted in \cite{tan94} that
the distribution of velocities
ceases to be Gaussian in the transition
regime between GOE and GUE.
The deviations are small, however,
and we have not been able
to reduce the statistical errors
of our numerical results
to an extent that a meaningful comparison
with the results of \cite{tan94} becomes possible.

\section{Deviations from RMT}
Within the non-linear sigma model it is possible
to derive $g^{-1}$--corrections to RMT.
For distributions of wave-function
amplitudes this was done in \cite{fyo95},
where in the orthogonal case one obtains
\begin{equation}\label{eq:co}
f(t)=
f^{\rm GOE}_{\mathrm PT}(t)\left[1+a_dg^{-1}\left(3/2-3t+
t^2/2\right)+{\cal O}\left(g^{-2}\right)\right]\,.
\end{equation}
In the case discussed in Ref. \cite{fyo95}, 
$a_3 \sim L/l$ where $L$ is
the linear dimension of the system
and $l$ is the mean free path.
Eq. (\ref{eq:co}) is valid provided $t \ll \sqrt{g/a_d}$.

We have performed numerical simulations in the $d=3$ Anderson model, 
using different values of disorder, and have 
computed the distribution $f(t)$ by
averaging over 400 realizations of disorder. The deviations from
the Porter--Thomas distribution, $\Delta f(t)=f(t)/
f^{\rm GOE}_{\mathrm PT}(t)-1$,
is shown in Fig.~\ref{fig:co}(b). 
In all cases, the deviations exhibit a
characteristic form: the probability of
finding small and large amplitudes
is enhanced, while the distribution
function is reduced near its maximum.
We find that $\Delta f(t)$ 
can be fitted using Eq.\ (\ref{eq:co}). At increasingly
large disorder, deviations occur at lower values of $t$.
In all cases, however, the zeroes of Eq.\ (\ref{eq:co})
are well reproduced.
Since at weak disorder  $l\sim W^{-2}$ one might expect
$a_3/g \sim W^4$. In the present case, however, we 
find $a_3/g \sim W^2$. In order to resolve this
discrepancy, more accurate numerical data at small
values of $W$ are needed. Corresponding results have been obtained 
for the unitary case.

\section{Conclusions}
We have analyzed RMT fluctuations of matrix elements and
level velocities in the $d=2$ dimensional Anderson model,
in the transition regime between GOE and GUE and have
found good agreement between the predictions of
RMT and our numerical results.
For the distribution of wave-function amplitudes
in $d=3$
we have studied deviations from RMT, in the
form of $g^{-1}$--corrections,
as suggested in \cite{fyo95}. Our numerical
results can be fitted by the expressions derived
in \cite{fyo95}, the dependence of
the fit parameter $a_3$  on $W$ however, differs 
from what might be expected. 
In this context it will be very interesting
to determine the tails of the distribution
functions and compare with
the predictions in \cite{fal95a,fal95b}
and \cite{smol97}.
    \vspace{0.6cm}\newline{\small 
Financial support by the DFG through
Sonderforschungsbereich 393 is gratefully acknowledged. 
V.U. thanks the DAAD for the financial support.
    }

\begin{thebibliography}{99}
\bibitem{note2} The condition
$\epsilon > \Delta$ ensures
that only periodic orbits
with periods $T_p < t_{\rm H} = 2\pi\hbar/\Delta$
contribute, so that the diagonal
approximation is valid
\end{thebibliography}
    \end{document}